\documentclass[twocolumn,amsmath,amssymb,nofootinbib,tighten,aps,pra]{revtex4-1}

\usepackage{graphicx}
\usepackage{dcolumn} 
\usepackage{bm}      
\usepackage{curves}
\usepackage{epic}
\usepackage{color}
\usepackage{epsf, times}
\usepackage{subfigure}
\usepackage{amsthm}

\def\openone{\leavevmode\hbox{\small1\kern-4.2pt\normalsize1}}

\def\P{\mathcal{P}}

\def\S{\mathcal{S}}

\def\tk{\tilde{k}}

\def\oa{\overline{\alpha}}
\def\ob{\overline{\beta}}
\def\erf{{\rm Erf}}

\newcommand{\beq}{\begin{equation}}
\newcommand{\eeq}{\end{equation}}
\newcommand{\bea}{\begin{eqnarray}}
\newcommand{\eea}{\end{eqnarray}}
\newcommand{\bfig}{\begin{figure}}
\newcommand{\efig}{\end{figure}}
\newcommand{\bei}{\begin{itemize}}
\newcommand{\eei}{\end{itemize}}

\begin{document}

\title{
Driven impurity in an ultracold 1D Bose gas with intermediate interaction strength
      }
    
\author{
Claudio Castelnovo$^{1,2,3}$ 
}
\author{
Jean-S\'{e}bastien Caux$^4$
}
\author{
Steven H. Simon$^1$
       }
\affiliation{
$^1$ 
Rudolf Peierls Centre for Theoretical Physics,
University of Oxford, 
1 Keble road, Oxford OX1 3NP, United Kingdom 
}
\affiliation{
$^2$ 
SEPnet and Hubbard Theory Consortium,
Department of Physics,
Royal Holloway University of London, 
Egham, TW20 0EX, United Kingdom
            }
\affiliation{
$^3$ 
TCM group,
Cavendish Laboratory,
University of Cambridge, 
Cambridge, CB3 0HE, United Kingdom
            }
\affiliation{
$^4$
Institute for Theoretical Physics, 
University of Amsterdam, 
1018 XE Amsterdam, The Netherlands
}

\date{\today}

\begin{abstract}

We study a single impurity driven by a constant force through a one-dimensional Bose gas using a Lieb-Liniger based approach. Our calculaton is exact in the interaction amongst the particles in the Bose gas, and is perturbative in the interaction between the gas and the impurity. In contrast to previous studies of this problem, we are able to handle arbitrary interaction strength for the Bose gas. We find very good agreement with recent experiments [Phys.~Rev.~Lett.~{\bf 103}, 150601 (2009)].   

\end{abstract}

\maketitle
%
%

\section{\label{sec: intro}
Introduction
        }
The study of quantum mechanical systems out of equilibrium is one of the great 
frontiers of modern physics. The questions in this field are
not only of fundamental interest, but are also of interest to future quantum 
technologies, as well as to classical technologies
on the nano-scale. Cold atomic systems have provided an ideal setting for  
hand-in-hand theoretical and experimental investigations of this frontier, 
particularly in low dimensions. Nonetheless, our understanding of the issues 
involved are sufficiently primitive that it remains useful to consider some 
of the simplest toy model experiments in order to gain intuition regarding 
more general questions. 

In this paper we focus on the study of driven impurities moving through a 
one-dimensional (1D) Bose gas. This subject has received much attention of late, 
thanks to both 
experimental~\cite{Palzer2009,Zipkes2010,Wicke2010,Catani2012,Fukuhara2013} 
and theoretical progress~\cite{Hakim1997,Zvonarev2007,Astrakharchik2004,%
Ponomarev2006,Girardeau2009,Sykes2009,Cherny2009,Gangardt2009,Johnson2011,%
Rutherford2011,Goold2011,Schecter2012,Bonart2012,Mathy2012,Knap2013}. 
Our work was inspired by the experimental results in 
Ref.~\onlinecite{Palzer2009}, where the impurity 
is driven through the gas by a constant force (gravity). 
This type of
experiment has not been 
extensively investigated at the 
theoretical level, with the exception of the recent work in 
Ref.~\onlinecite{Rutherford2011}  (see also Ref.~\onlinecite{Knap2013}) 
which uses a Tonks-Gas description~\cite{Girardeau1960} 
(appropriate only in the limit of 
large interaction strength). We also refer the reader to 
Ref.~\onlinecite{Johnson2011}, where a similar system was studied in 
presence of a 1D optical lattice. 

In this work, we use linear response theory and Fermi's golden rule with 
exact transition rates to 
model the scattering between the driven impurity and the underlying gas 
(see e.g., Refs.~\onlinecite{Palzer2009}, \onlinecite{Johnson2011}, 
\onlinecite{Cherny2012}). 
We then model the motion of the impurity as a classical driven stochastic 
process~\cite{footnote: approach}. Our approach is strictly valid in the limit
where the interaction between the impurity and the underlying gas is 
sufficiently weak (which is not necessarily true in the experiments of 
Ref.~\onlinecite{Palzer2009}). However, in contrast to prior works attempting 
to analyze this problem, 
our approach is valid for any interaction strength between particles in the 
1D gas.  
The main point of this work is to provide a method to analyze 
the effects of interaction within the 1D gas on a driven impurity -- which 
has previously not been possible for 
intermediate interaction strength within the gas. 

We find quantitative agreement with the results in  
Ref.~\onlinecite{Palzer2009}
both comparing the center of mass motion as 
well as the optical density profile of a packet of impurities after they 
leave the 1D gas. In contrast to  
earlier theoretical work~\cite{Rutherford2011,Knap2013} which found agreement 
in the strong interaction limit 
with Tonks-Girardeau modelling, our results quantitatively describe the 
experimental data for small 
values of the dimensionless interaction strength $\gamma \sim 1-3$, 
close to the Bogoliubov limit~\cite{Bogoliubov1947}, 
in keeping with the experimental 
parameter range (see Table~\ref{tab: gamma}).  
%
%

\section{\label{sec: method}
Method
        }
Our general method for handling driven impurtity motion is to treat the 
scattering of the impurity atom perturbatively.
Our second key assumption is that the driven impurity atom is a negligible 
perturbation of the underlying 1D system, which is assumed to relax to its 
ground state between any two scattering events. 
This assumption is strictly valid when the impurity scatters only once 
in the entire time span of the experiment (which is approximately the case 
if the coupling of the impurity to the 1D gas is weak). 
Nonetheless, there are 
several other regimes for which this approximation is expected to be quite 
good. For example, the behaviour of the 1D gas may not differ much
if it is in its exact ground state versus being slightly excited.  Another 
regime of interest is when the impurtity moves faster than the effective speed 
of propagation in the 1D gas.  In this case if the impurity scatters a second 
time, it will have out-run the perturbation it caused in the first scattering 
event and will effectively see the 1D gas as if it were in its ground 
state. 

We consider a delta function interaction potential of interaction strength 
$g_{im}$, $\hat{V}=g_{im} \sum_i \delta(x - x_i)$, 
between the driven impurity atom at position $x$ and the 1D gas atoms at 
positions $\{ x_i \}$. 
In the usual way, Fermi's golden rule gives a transition rate between an 
initial $|i\rangle$ and final $|f\rangle$ state of the system, of energies 
$E^0_i$ and $E^0_f$ respectively, as 
\bea
W_{\rm if} 
= 
\frac{2 \pi}{\hbar} 
\left\vert 
  \langle f^0 \vert \hat{V} \vert i^0 \rangle 
\right\vert^2
\delta\left( E^0_f - E^0_i \right)
, 
\eea
where the superscript ${}^0$ indicates that these states and energies are to 
be evaluated in the absence of the coupling 
$\hat V$ between the impurity and the gas.  Hence we have  $\vert i^0 \rangle$, 
$\vert f^0 \rangle = \vert k \rangle \otimes \vert n \rangle$, 
where $H_{\rm 1D\,gas} \vert n \rangle = \varepsilon_n \vert n \rangle$ and 
$H_{\rm impurity} \vert k \rangle = (\hbar^2 k^2 / 2 m_{im}) \vert k \rangle$ 
with $m_{im}$ the impurity mass, $H_{\rm impurity}$ the Hamiltonian of the 
impurity, $H_{\rm 1D gas}$ and $\varepsilon_n$ the Hamiltonian and 
eigenenergies of the 1D gas. 

As described above, we assume that only the ground state $n=0$ appears in 
$\vert i^0 \rangle$. Summing over all final states of the 1D gas, we obtain 
a transition rate for the impurity
\bea
W_{k \to k^\prime} 
\!&=&\! 
\frac{g_{im}^2}{\hbar L} 
\sum_n 
\left\vert\vphantom{\sum} 
  \langle n \vert \rho_{k^\prime-k} \vert 0 \rangle 
\right\vert^2
\!\delta\!\left[ 
  \varepsilon_n - \varepsilon_0 + 
  \frac{\hbar^2 ({k^\prime}^2 - k^2) }{2m_{im}} 
\right]
\nonumber \\ 
&=& 
\frac{g_{im}^2}{\hbar L} 
\frac{N_p}{\varepsilon_F}
\:\S\!\left( 
  k-k^\prime, \frac{\hbar^2 (k^2 - {k^\prime}^2)}{2m_{im}} 
\right)
, 
\label{eq: W k to k prime}
\eea
where $\rho_{k^\prime-k} \equiv \sum_i e^{-i (k^\prime-k) x_i}$ is the 
Fourier transform of the density operator of the 
1D gas and $\S(k,\omega)$ is the \emph{dimensionless} dynamic 
structure factor 
(DSF) of the 1D gas (note the factor of $\varepsilon_F / N_p$ in our 
definition of $\S$, $N_p$ being the number of particles and 
$\varepsilon_F \equiv \hbar^2 k_F^2 / 2 m$ their Fermi energy; 
here $m$ is the mass of the particles in the 1D gas and $k_F \equiv \pi n$ 
is the Fermi wavevector with $n=N_p/L$ the density).

We assume our 1D gas is made of spinless bosons and has short ranged 
interactions of the form $g_{1D} \sum_{i < j} \delta(x_i - x_j)$. 
For convenience, we introduce the standard dimensionless 
interaction strength $\gamma \equiv g_{1D} m / (\hbar^2 n)$.
Analytical solutions for $\S(k,\omega)$ are available in the weakly  or 
strongly interacting limit. 
For intermediate values of $\gamma$  one may use the exact Lieb-Liniger 
(LL) solution for the DSF which can be obtained numerically for any values 
of $k$ and $\omega$.  A description of this numerical procedure can be found 
in Ref.~\onlinecite{Caux}. 

Once we can calculate the transition rate, we need to account for the driven motion of the impurity. To simulate both the driving force and the scattering, we discretize time and momentum and write a scattering transition probability $\P_{k \to k^\prime} \equiv W_{k \to k^\prime} \delta t \Delta k$ and we define the probability of not scattering to be $\P_{k \to k} \equiv 1 - \sum_{k^\prime \neq k} \P_{k \to k^\prime}$. For every time interval $\delta t$, we evolve the position and velocity of the particles deterministically.   In the present case (inspired by the experiments of Ref.~\onlinecite{Palzer2009}) we are concerned with the impurity being accelerated (driven) by gravity $g$ (assume acceleration in the $+x$ direction) so we have
\beq
\begin{array}{rl}
& x(t+\delta t) = x(t) + v(t) \delta t + g \delta t^2 / 2 
\\
& v(t+\delta t) = v(t) + g \delta t. 
\end{array}
\eeq
After each time interval $\delta t$ we  then allow for a stochastic 
scattering attempt $v_1 \equiv v(t+\delta t) \to v_2$ with probability per 
unit wave vector $\P_{v_1 \to v_2} = W_{k \to k^\prime} \,\delta t$, 
where $v_1 = \hbar k / m_{im}$ and $v_2 = \hbar k^\prime / m_{im}$. 
This allows for efficient 
simulation of the impurity motion. 
In the large and small $\gamma$ regime, we have used analytic forms of the 
DSF to test our numerical algorithm. 
%
%

\section{\label{sec: expm}
Experimental Parameters
        } 
As demonstration of our method we apply it 
to the experimental situation from Ref.~\onlinecite{Palzer2009}. A Bose condensate of $^{87}$Rb atoms is confined into an ensemble 
of  harmonic traps with long axis aligned with the Earth gravitational field. 
The transverse radius of each trap and temperature 
are such that each system is in an effective zero-temperature 1D regime.
The parameters of the 1D traps vary with position 
-- both between different 1D systems in the ensemble and within each 
individual 1D system. 
It should be noted that the value of $\gamma$ is 
expected to vary significantly across the system. 
Private communication with the authors of Ref.~\onlinecite{Palzer2009} lead 
to the estimates reported in Table~\ref{tab: gamma}. 
\begin{table}
\begin{tabular}{c|c|c|c}
parameters & (1) & (2) & (3)
\\
\hline
$N_p$ in central 1D system & 36 & 32 & 30 
\\
$L$ of central 1D system ($\mu$m) & 22.46 & 24.1 & 24.82 
\\
$\gamma_{\rm ave}$ average over entire condensate & 3 & 5 & 7 
\\
$\gamma_{\rm ct}$ average over central 1D system & 1.7 & 2.9 & 4.0 
\\
$\gamma_{\rm c}$ at the center of condensate & 1.1 & 1.9 & 2.6 
\end{tabular}
\caption{
\label{tab: gamma}
Range of parameters of the 1D systems obtained from three different Bose 
condensates (columns 1-3)~\cite{SiasKoehl}. 
}
\end{table}

For simplicity we crudely neglect the nonuniformity of the system, considering 
only the case of a homogeneous 1D system with fixed 
density~\cite{footnote: 1D gas profile}. 

A radio-frequency (RF) pulse is used to change the hyperfine ground state of 
some (up to 3) atoms near the center of the 1D system so as to decouple them 
from the trap. 
The pulse is Fourier-limited in width [full width at half maximum (FWHM) 
$\sim 2.3$~$\mu$m] and has a 
velocity distribution of width $\simeq 2\cdot10^{-3}$~m/s (close to the 
uncertainty limit). 
Therefore, we consider a wave packet 
$\psi(x) \propto \sin \left( \alpha x \right) / x$, 
where $\alpha = 2 \pi/\Delta x \simeq 2.73$~$\mu$m$^{-1}$ 
(although we find that $\alpha \simeq 1.8-2.0$~$\mu$m$^{-1}$ 
produces a better fit to the width of the unscattered peak in the 
experimental density profile of the falling atoms at long times, 
illustrated in Fig.~\ref{fig: xv hist ft}). 

We model these initial conditions using a 
Gaussian-smoothed~\cite{footnote: smoothing} 
Wigner function~\cite{Cartwright1976} for the position and momentum 
distribution of the falling atoms at time $t=0$: 
\bea
G(x,p) 
&\propto& 
\int_{- \infty}^{\infty} 
  W(x',p') 
  e^{- \frac{\oa (x - x')^2}{\hbar}}
	e^{- \frac{\ob (p - p')^2}{\hbar}}
\: dx' dp' 
\nonumber \\ 
W(x,p) &\propto& 
\int_{- \infty}^{\infty} 
  \psi^*(x+y) \psi(x-y) e^{2 i p y / \hbar}
\: dy 
, 
\eea
where $\oa$ and $\ob$ are positive real constants that satisfy the condition 
$\oa \ob \leq 1$, i.e., the smoothing area is $\geq \hbar$. We 
choose the least possible smoothing that yields a positive semi-definite 
probability, namely $\oa \ob = 1$. 
The value of $\oa$ is then set to equal $\hbar \alpha^2$. 
After a few lines of algebra one obtains: 
\bea
G(x,p) 
&\propto& 
e^{- \alpha^2 x^2} 
\left\vert 
  \erf\left(\frac{1 + \frac{p}{\hbar\alpha} + i \alpha x}
	               {\sqrt{2}}\right)
\right. 
\nonumber \\ 
&& 
\qquad \;\;
\left.
	+
	\erf\left(\frac{1 - \frac{p}{\hbar\alpha} - i \alpha x}
	               {\sqrt{2}}\right)
\right\vert^2 
,
\label{eq:Gfinal}
\eea
where $\erf$ is the Gaussian error function extended to the complex plane. 
This distribution was sampled using the rejection sampling technique. 

The decoupled impurity atoms are then allowed to accelerate under the 
constant driving force of the gravitational field. In our simulations we 
assume only a single impurity atom is decoupled from the trap (i.e, 
we neglect interactions between multiple falling impurity atoms; 
we also disregard possible effective interactions between the decoupled 
atoms that may be mediated by the condensate). 

In this experiment, the falling (impurity) atoms are identical to the atoms 
in the trap (gas) up to their spin state. Hence, $m_{im} = m$ and all 
interactions (impurity-gas and gas-gas) are described by the same 
delta function potential, $g_{im}=g_{1D}$. 
%
%

\section{\label{sec: com}
Center of Mass
        } 
We start by considering the position of the overall center of mass of the 
packet of falling atoms as a function of time, which was measured 
experimentally and reported in Fig.~3 of Ref.~\onlinecite{Palzer2009}. 
The parameters used in this experiment are those listed as case (3) in 
Table~\ref{tab: gamma}. 

\begin{figure}[ht]
\includegraphics[width=0.95\columnwidth]
                {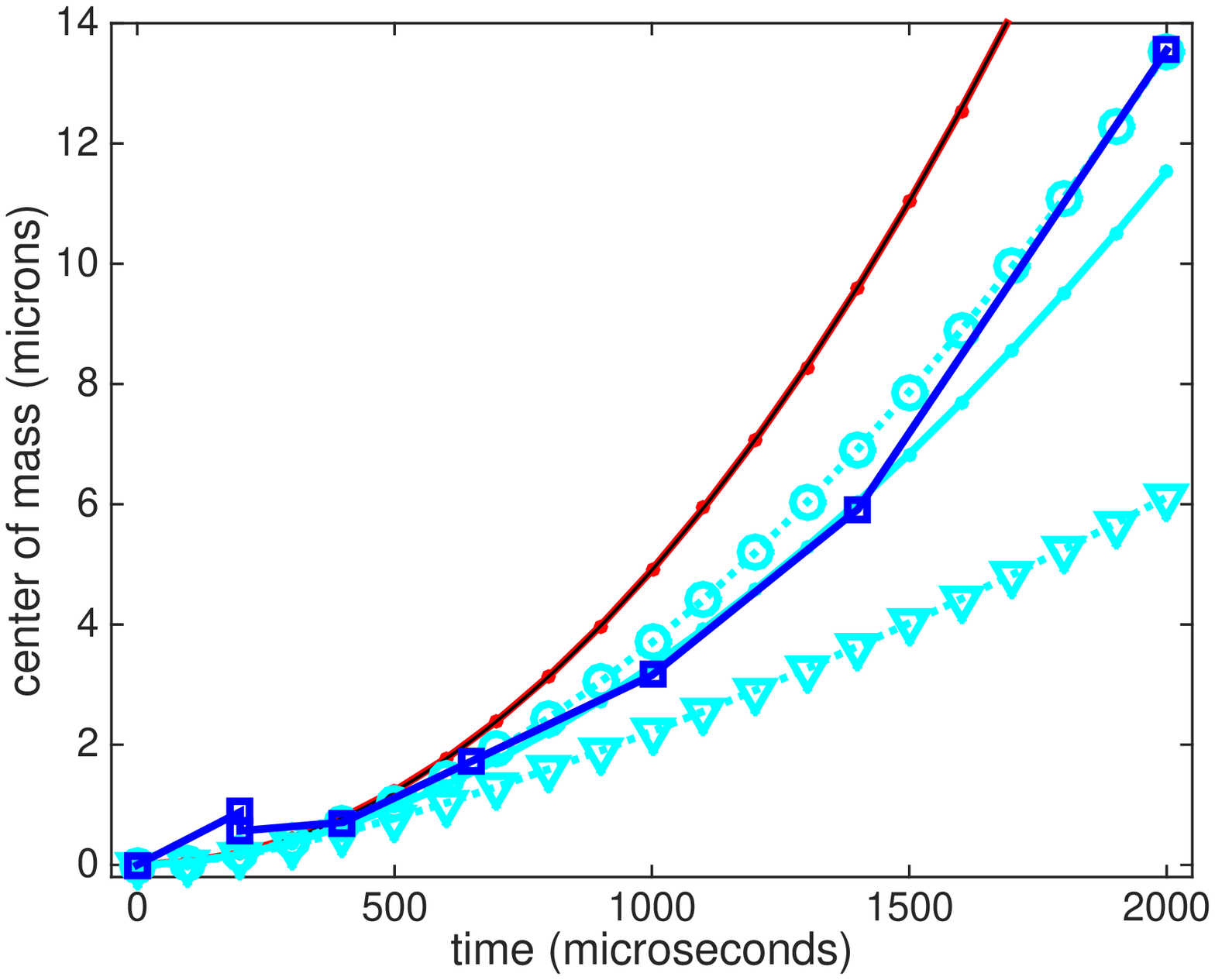}
\\
\includegraphics[width=0.95\columnwidth]
                {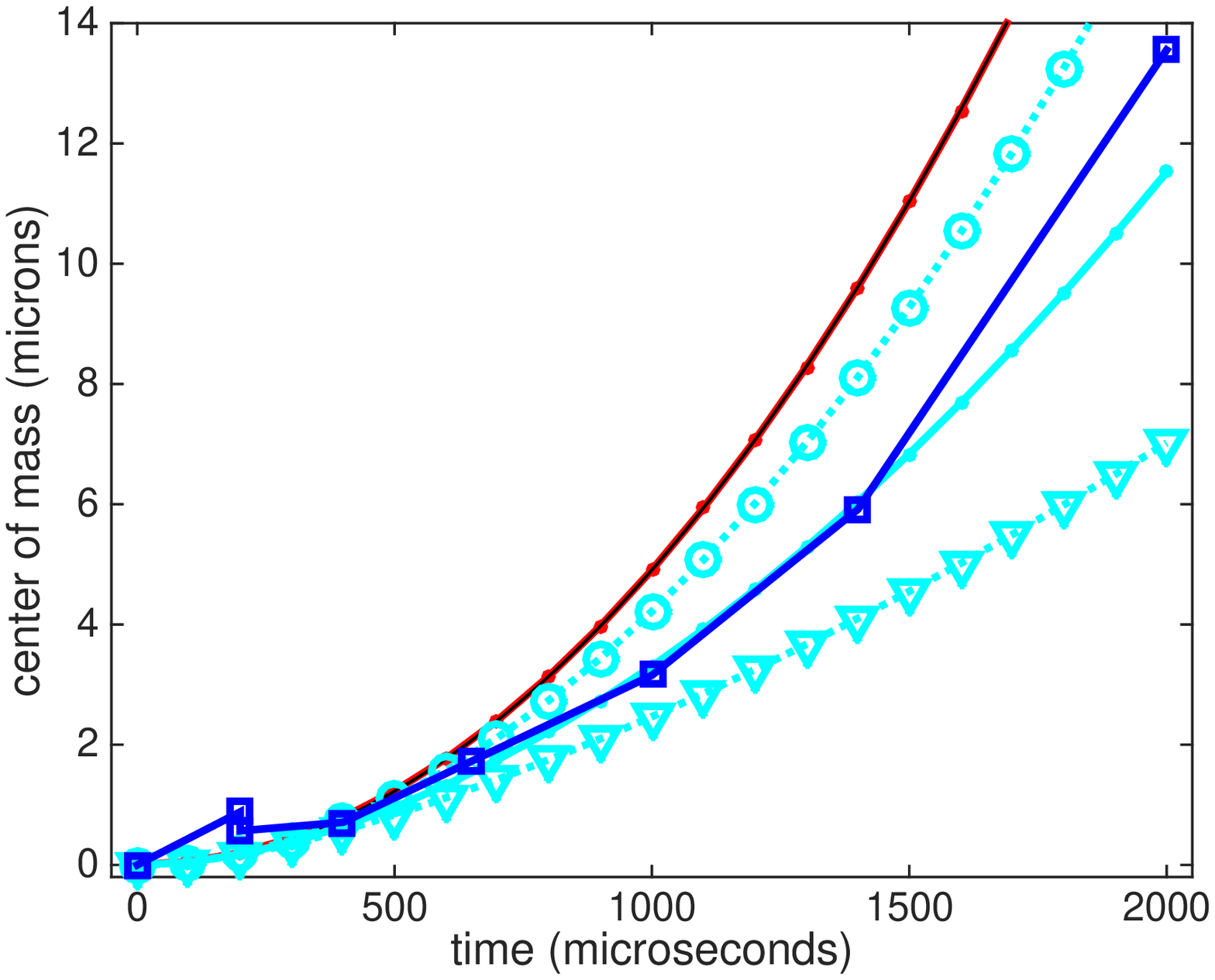}
\caption{
\label{fig: cm vs expm} 
(Colour online) --- 
Position of the center of mass as a function of time from the numerical 
simulations in comparison with the experimental data from Fig.~3 in 
Ref.~\onlinecite{Palzer2009} (blue open squares). 
The free-fall analytic solution $x(t) = g t^2 / 2$ is also shown for 
comparison (red solid dots). 
Top panel: numerical results at fixed density 
$n \simeq 1.2$~$\mu$m$^{-1}$, for 
$\gamma = 1.9$ (cyan open circles), 
$\gamma = 2.6$ (cyan solid dots), and 
$\gamma = 4.0$ (cyan open triangles). 
Bottom panel: numerical results at fixed $\gamma = 2.6$, for 
$n \simeq 0.84$~$\mu$m$^{-1}$ (cyan open circles), 
$n \simeq 1.2$~$\mu$m$^{-1}$(cyan solid dots), and 
$n \simeq 1.56$~$\mu$m$^{-1}$ (cyan open triangles). 
}
\end{figure}

Fig.~\ref{fig: cm vs expm} shows a comparison between the experimental 
curves from Ref.~\onlinecite{Palzer2009} and the results from our stochatic 
simulations, using $n = N_p/L \simeq 1.2$~$\mu$m$^{-1}$ and different values 
of $\gamma$ (top panel), as well as using $\gamma = 2.6$ and different 
values of $n$ (bottom panel). 
In the simulations we consider an infinite 1D gas of uniform density. 
The experimental time $t=0$ in Ref.~\onlinecite{Palzer2009} was chosen to 
correspond to the middle of the RF pulse that creates the packet of falling 
atoms. Accordingly, we chose $t=0$ in our simulations as the time when the 
Fourier limited packet starts moving through the 1D gas. 

The numerical results appear to be very sensitive to the values of $\gamma$ 
and $n$. This allows us to determine that the combination 
$n \simeq 1.2$~$\mu$m$^{-1}$ and $\gamma = 2.6$ provides the best fit to 
the experimental data. 
(We note that particles begin to leave the 1D trapped gas after about $2$~ms, 
which corresponds to the longest time reported in Fig.~\ref{fig: cm vs expm}. 
Such an effect may be responsible for the discrepancy that we observe between 
our numerics and the very last data point in the experiment.) 

Our results are in contrast with earlier theoretical 
modelling~\cite{Palzer2009,Rutherford2011,Knap2013} which achieved a 
similarly good fit to the experimental results by using the strongly 
interacting Tonks-Girardeau (TG) approximation~\cite{Girardeau1960} 
corresponding to $\gamma = \infty$ within the 1D gas and then treating 
the interaction between the impurities and the 1D gas at mean field level 
with an intermediate interaction strength $\gamma=7$ 
(see also Appendix.~\ref{app: TG limit}). 
%
%

\section{\label{sec: dens profile}
Density Profile
        } 
In order to further test our approach, we computed 
the profile of the falling atoms at long times after they exit the 
1D gas, which can be compared with the experimental results reported 
in Fig.~5 of Ref.~\onlinecite{Palzer2009}. 
Experimental results are available~\cite{Palzer2009} for all three cases 
in Table~\ref{tab: gamma}. 
Unfortunately, a similar comparison was not carried out in earlier theoretical 
modelling~\cite{Palzer2009,Rutherford2011,Knap2013}. 

In our simulations, we approximate the 1D gas to have uniform density and 
fixed length, with parameters $N_p$ and $L$ as in the experiments. 
Once again, we find that the resulting density profile of the falling packet 
has a significant sensitivity on the value of $\gamma$, which allows us to 
readily identify which one gives the best fit. 

\begin{figure}[ht]
\includegraphics[width=\columnwidth]
                {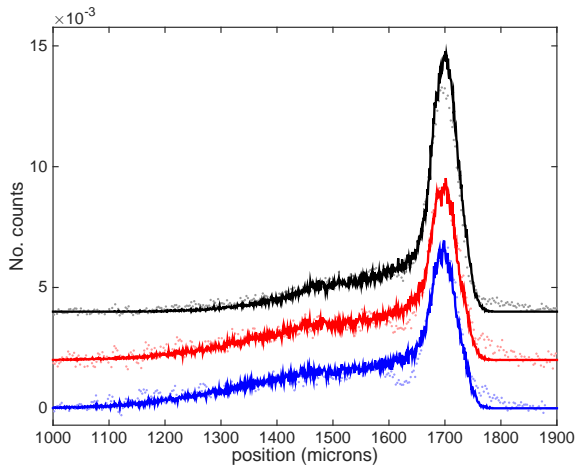}
\caption{
\label{fig: xv hist ft} 
Behaviour starting from gaussian distributed initial conditions over 
$N_{\rm hist} = 100,000$ histories, for $\gamma = 1.1, 1.9, 2.1$ 
(black, red and blue, respectively) with a finite trap of uniform density 
(see text for size and density details). 
The results are expressed as histograms of the position of the particles 
after they have been falling for 
$18.6$~ms~\cite{footnote: expm corrections long time}. 
The corresponding experimental results in Fig.~5 of 
Ref.~\onlinecite{Palzer2009} are shown as thin dotted lines. 
}
\end{figure}

The outcome is shown in Fig.~\ref{fig: xv hist ft}. 
The experimental and simulation curves were normalised so that the area 
under the profiles equals $1$ (after subtraction of 
a background~\cite{footnote: expm corrections background}). 
The main peak in the figure is due to the fraction of particles 
that fall freely through the 1D gas without 
scattering~\cite{footnote: expm corrections long time}. 

Both the overall shape of the curves and the ratio between scattered and 
free-falling contributions are in reasonable agreement between 
numerics and experiments for $\gamma = 1.1,\, 1.9,\,{\rm and}\, 2.1$, 
respectively. These results suggest that the relevant 
values of $\gamma$ in the experiments are those from the central 
region of the condensate. 

We note that for these values of $\gamma$ we find very good agreement 
between the exact LL solution and the Bose gas (BG) 
approximation~\cite{Bogoliubov1947}. 
In the BG limit, we studied also 1D gases with static position-dependent 
density~\cite{footnote: 1D gas profile}. 
We found that the resulting effects are minor and do not alter the best fit 
values of $\gamma$. 

We notice that a small dip between the scattered peak and the 
free-falling peak appears in the experimental data (most noticeably at 
larger values of $\gamma$) whereas it is nearly absent in the numerical 
simulations. We conjecture that this dip might be due to the fact that 
in the actual experiment two or more impurities might fall though the trap at 
the same time. An effective attractive interaction between impurities could 
bind together nearby impurities and enhance the main peak at the expense 
of weight on either sides of the main peak. 
This effect is beyond our approximation and must be relegated to future 
research. 
%
%

\section{\label{sec: conclusions}
Conclusions
        }
Using linear response theory and Fermi's golden rule with exact transition 
rates to model the scattering 
between the driven impurity and the underlying 1D Bose gas, we have been able 
to obtain a quantitative description of the experimental results 
in Ref.~\onlinecite{Palzer2009}: 
center of mass, profile of driven packet with time-of-flight measurements 
and tomography. It appears that our crude approach is entirely sufficient to 
describe the vast majority of the observed physics. 
If desired, the approach taken here could be systematically improved by 
considering corrections of higher order in the coupling between the Bose gas 
and the impurity. 
The detection of finer quantum mechanical effects beyond our description 
may however require higher experimental accuracy. 

In the range of $\gamma$ values considered here, the need for an exact LL 
solution was limited and the results would have been in large part the same 
had we used the BG approximation instead. 
However, sizeable differences between LL and BG arise already for 
$\gamma \gtrsim 3$, which is within experimental reach. 

The case considered here can be viewed as an extremely simple example of 
handling a non-equilibrium situation in the presence of strong correlations. 
While the impurity is driven, the physics of the Bose gas can still be 
understood as remaining at equilibrium. 
Going further, one could try to understand how putting the gas itself out of 
equilibrium affects the impurity dynamics. 
Moreover, besides cold atom settings, one could also consider driven quantum 
magnets, for which the necessary exact correlators are also available. 
We will return to these issues in future work. 
%
%

\begin{acknowledgments}
The authors are very grateful to M.~K\"{o}hl and C.~Sias for helpful 
discussions and particularly for sharing generously detailed information about
the experiments of Ref.~\onlinecite{Palzer2009}. The authors also thank 
D.~Kovrizhin and Z.~Hadzibabic for helpful discussions. 
This work was supported in part by 
EPSRC Grants EP/G049394/1 and EP/K028960/1 (CC), EP/I032487/1 and 
EP/I031014/1 (SS), and by the FOM and NWO foundations of the Netherlands (JSC). 
Statement of compliance with EPSRC policy framework on research data: 
This publication reports theoretical work that does not require supporting 
research data. 
\end{acknowledgments}
%
%
\appendix
%
%

\section{\label{app: 1D gas profile dep}
Dependence on 1D gas density profile
        }
For the values of $\gamma$ of relevance to the experiments in 
Ref.~\onlinecite{Palzer2009}, we find that our simulations give similar results 
whether we use the DSF from LL or in the BG approximation. 
We can therefore use the latter to test how the results are affected by 
(static) changes in the 1D gas density profile. 

The DSF of a 1D bose gas can be determined directly from
its spectrum~\cite{Bogoliubov1947} 
\bea
\hbar \omega_k 
&=& 
\frac{\hbar^2 k^2}{2m} 
\sqrt{1 + \frac{4 \gamma k_F^2}{\pi^2 k^2}} 
, 
\label{eq: bogoliubov spectrum}
\eea
using the f-sum rule: 
\bea
S(k,\omega) 
&=& 
\frac{N_p}{\hbar} 
\left[
  1 
  + 
  \frac{4\gamma}{\pi^2 \left( k/k_F \right)^2} 
\right]^{-1/2}
\!\!\!\!\!\!
\delta(\omega - \omega_k) 
. 
\label{eq: boson gas structure factor I}
\eea
After a few lines of algebra, following the steps outlined in 
Sec.~\ref{sec: method}, one obtains that the only allowed outgoing wave 
vector is $k^\prime = \gamma k_F^2 / (\pi^2 k)$, with probability 
\bea
\P_{k \to \gamma k_F^2 / \pi^2 k} 
&=& 
\left\{
\begin{array}{lcl}
\frac{2 \gamma^2}{\pi^3} 
\frac{\varepsilon_F}{\hbar} 
\delta t 
\frac{k_F}{\vert k \vert}
& 
\qquad
&
\vert k \vert / k_F > \sqrt{\gamma} / \pi
\\ 
0 
& 
\qquad
& 
{\rm otherwise.} 
\end{array}
\right.
\label{appeq: boson gas scattering probability} 
\eea
Notice that Eq.~\eqref{appeq: boson gas scattering probability} can be 
interpreted as a probability only if it is $\leq 1$, which in turn is 
satisfied if we choose 
\bea
\delta t 
\leq 
\frac{\pi^2}{2 \gamma^{3/2}}
\frac{\hbar}{\varepsilon_F} 
, 
\eea
where we used explicitly the condition 
$\vert k \vert / k_F > \sqrt{\gamma} / \pi$. 

\begin{figure}[ht]
\includegraphics[width=0.95\columnwidth,height=0.55\columnwidth]
                {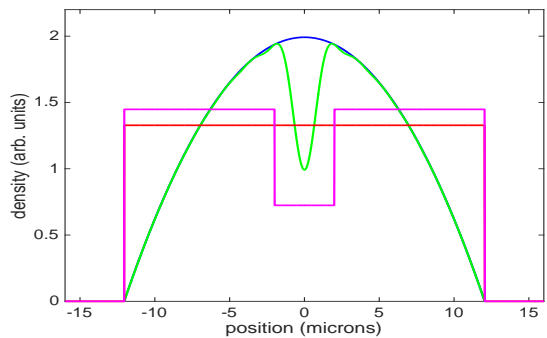}
\caption{
\label{fig: condensate shapes BG} 
Different (static) density profiles for the 1D condensate used to assess 
how the shape affects the results of our simulations: 
(i) square (red), 
(ii) square with a square depletion at the center of the condensate (magenta), 
(iii) parabolic (blue), and 
(iv) parabolic with a Gaussian-smoothed Wigner function depletion at the 
center of the condensate (green). 
}
\end{figure}

Using Eq.~\eqref{appeq: boson gas scattering probability} one can 
straightforwardly adapt the simulations to a position-dependent (static) 
density profile of the underlying 1D gas. 
For concreteness we fix the average density at the experimentally relevant 
value of $1.3278$ particles per micrometre (corresponding, in the case of 
uniform density, to an average $\gamma = 1.9$). 
We then contrast the following cases: 
(i) a uniform condensate of finite length $L = 24.1$~$\mu$m; 
(ii) a uniform condensate of the same length 
with a square depletion to half its density near its center 
(defined as $-2.0 < x < 2.0$~$\mu$m); 
(iii) a parabolic condensate of the same length and average density; 
and finally
(iv) a parabolic condensate with a central depletion obtained by subtracting 
the Gaussian-smoothed Wigner function that we used to describe the initial 
distribution of the falling packet, Eq.~\eqref{eq:Gfinal}, after setting $p=0$. 
The four different options are illustrated in 
Fig.~\ref{fig: condensate shapes BG}. 
The depleted cases are intended to mimic the effect of the decoupling laser 
that excites some of the atoms in the condensate to a non-trapped state, thus 
creating the initial packet of falling atoms (cf. Fig.~2 in 
Ref.~\onlinecite{Palzer2009}). Note that we continue to neglect 
any feedback between the falling atoms and the condensate, nor we 
allow the latter to relax to a shape different form the initial one. 

\begin{figure*}
\includegraphics[width=0.32\textwidth]
                {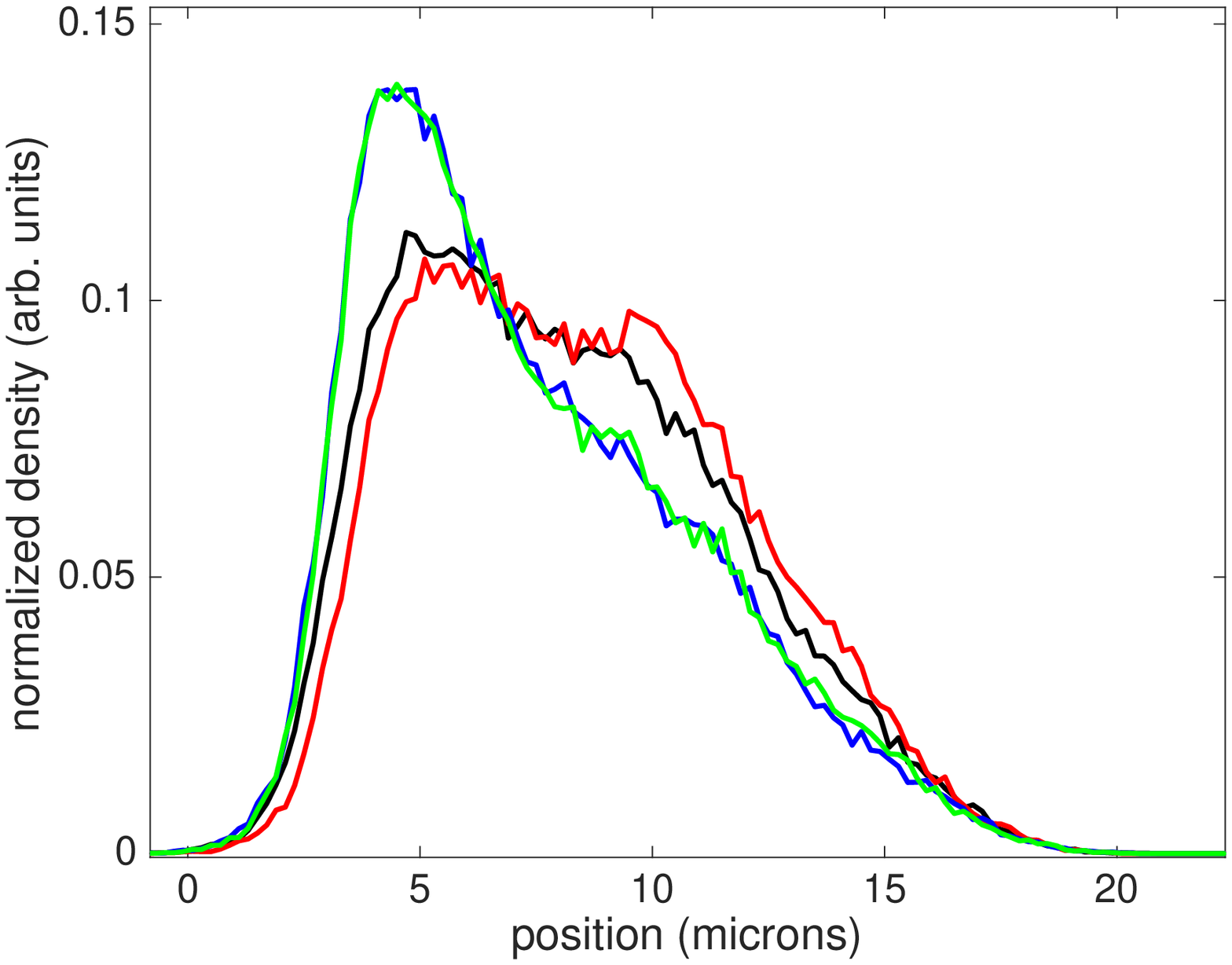}
\includegraphics[width=0.32\textwidth]
                {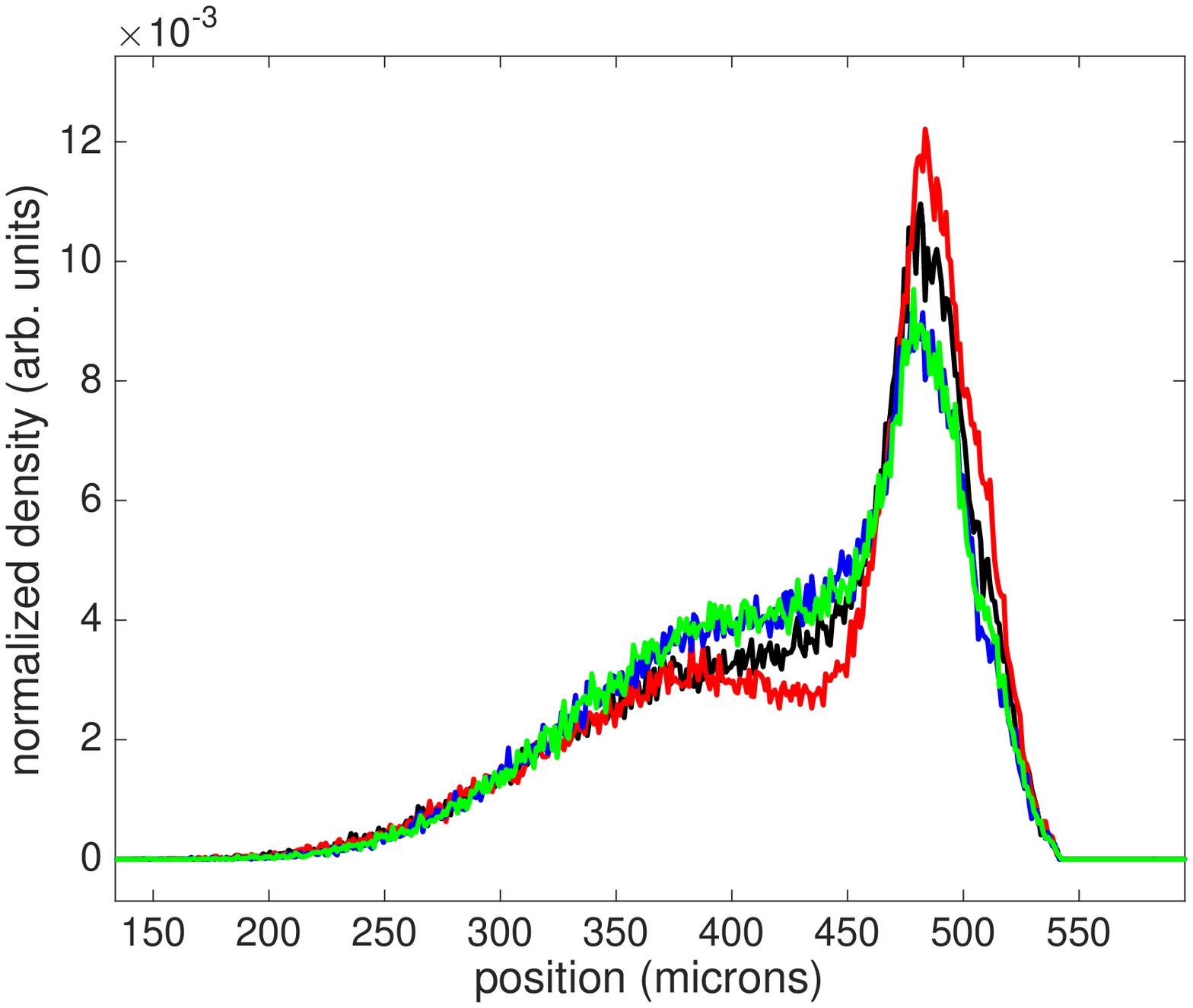}
\includegraphics[width=0.32\textwidth]
                {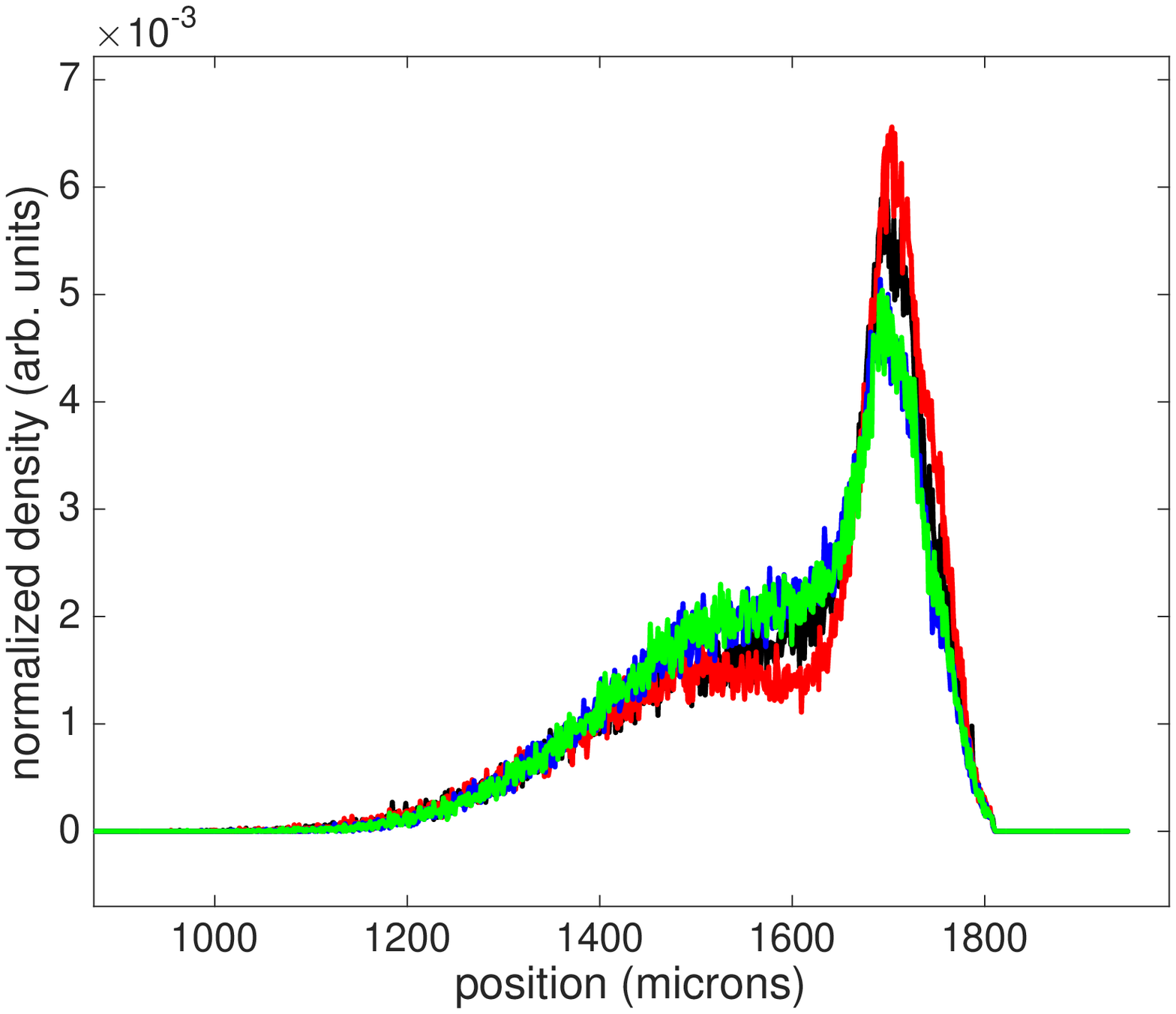}
\caption{
\label{fig: bg results} 
Density profiles at different times starting from the initial 
conditions discussed in the main text, averaged over $N_{\rm hist} = 100,000$ 
histories. 
From left to right: $t = 1.5$~ms, $t = 10$~ms, and $t = 18.7$~ms. 
The colours correspond to the four cases discussed in the text: 
(i) a finite uniform condensate (black); 
(ii) a finite uniform condensate with a square depletion at the center (red); 
(iii) a parabolic condensate (blue); and 
(iv) a parabolic condensate with a Gaussian-smoothed Wigner function 
depletion at the center (green). 
}
\end{figure*}

Fig.~\ref{fig: bg results} shows the density profiles of the falling atoms 
at different times, using the same initial conditions 
discussed in the main text. 
The differences are minor and comparable to the 
experimental error bars in Ref.~\onlinecite{Palzer2009}. 
The case of a parabolic profile ought to be considered with care, since a 
continuously vanishing density at its edges implies large values of $\gamma$, 
and the BG approximation is no longer justified. 

We notice that Ref.~\onlinecite{Johnson2011}, which considers a similar 
system in presence of a 1D optical lattice, also reported qualitatively 
similar results whether the 1D gas was prepared in equilibrium with or 
without the impurities (see the third paragraph in Sec.II D of 
Ref.~\onlinecite{Johnson2011}). 
%
%

\section{\label{app: TG limit}
Tonks-Girardeau limit
        }
In our work we have found quantitative agreement with the experimental 
results in Ref.~\onlinecite{Palzer2009} for small values of $\gamma$ where 
the BG approximation is reasonably accurate. 
This is in contrast with the modelling presented in that very same 
reference~\cite{Palzer2009}, as well as the work done in 
Ref.~\onlinecite{Rutherford2011} and Ref.~\onlinecite{Knap2013}, 
which make use of the Tonks-Girardeau (TG) limit. 

In this section we investigate the motion of the center of mass of the 
falling packet in the TG limit using our method. A reasonable agreement with 
the experimental results can be obtained only in the small $\gamma$ limit, 
which is in contradiction with the TG approximation. 
According to our simulations, already at intermediate values of the 
coupling strength (namely, $\gamma \gtrsim 7$) 
the falling atoms reach terminal velocity well within the time of the 
experiment, in contrast with the observed behaviour. 

The dynamic structure factor of a 1D gas in the TG 
limit~\cite{Girardeau1960} ($\gamma\to\infty$) can be written as 
\bea
S(k,\omega) 
&=& 
\frac{N_p}{4 \varepsilon_F}\frac{k_F}{k} 
\left[ 
  \Theta(\omega - \omega_{-}) \Theta(\omega_{+} - \omega) 
\right. 
\nonumber \\ 
&& 
\qquad\;\;\,
- 
\left.
  \Theta(\omega - \omega_{+}) \Theta(\omega_{-} - \omega) 
\right]
, 
\label{eq: tonks gas structure factor I}
\eea
where 
\bea
\omega_{\pm}(k) 
= 
\frac{\hbar {k_F}^2}{2m} 
  \left\vert 2\frac{k}{k_F} \pm \frac{k^2}{k_F^2} \right\vert
. 
\label{eq: tonks gas structure factor -- omega}
\eea
Introducing the dimensionless wave vector notation $\tk = k/k_F$, after 
the usual substitution $k = k^\prime - k$ and 
$\omega = \hbar^2 {k^\prime}^2 / 2m - \hbar^2 k^2 / 2m$, 
a few lines of algebra show that the scattering probability density per unit 
of dimensionless wave vector, from $k$ to $k^\prime$, is given by 
\bea
\P_{k \to k^\prime} 
&=& 
\begin{cases}
  \frac{\gamma^2}{\pi^3} 
  \frac{\varepsilon_F}{\hbar} 
  \delta t 
  \frac{1}{\vert \tk^\prime - \tk \vert}
  & \textrm{if}\;\;\vert \tk \vert > 1\;\;
    \textrm{and}\;\;\vert \tk^\prime \vert < 1
  \\
  0 & \textrm{otherwise.} 
\end{cases}
\label{eq: TG probability}
\eea

The expression above, which is correct to leading order in $\gamma$, 
presents the intrinsic problem that the total scattering probability at a 
given time, 
\bea
\int^{+\infty}_{-\infty} \P_{k \to k^\prime} \: d\tk^\prime 
= 
\frac{\gamma^2}{\pi^3} \frac{\varepsilon_F}{\hbar} \delta t 
  \left\lvert \ln \left( \frac{\tk-1}{\tk+1} \right) \right\vert
, 
\eea
diverges in the limit $k \to k_F$. For the stochastic approach to be valid, 
a necessary condition is that $\delta t$ be small enough so 
that the integrated probability at any given time remains 
smaller than $1$, which thus requires $\delta t$ to be vanishingly small 
for $k$ arbitrarily close to $k_F$. 

The singularity is directly related to the limit $\gamma \to \infty$. 
However, it cannot be easily resolved by including the subleading 
correction in $1/\gamma$ because the expansion of $\mathcal{S}$ 
becomes negative in some range of $k$ and $\omega$~\cite{BrandCherny}. 

A compromise to obtain a non-negative, non-divergent probability is to 
use the expansion of $\mathcal{S}$ to leading order, as in 
Eq.~\eqref{eq: tonks gas structure factor I}, but to replace the Heaviside 
Theta functions with those from the Random Phase Approximation (RPA). 
Namely, we use the TG form of the DSF, but with support in $k$ and $\omega$ 
from RPA. This in turn means that the probability $\P_{k \to k^\prime}$ 
retains the same form as in Eq.~\eqref{eq: TG probability}, but it is 
set to zero identically outside the range: 
\bea
\begin{cases}
1-\frac{4}{\gamma} < \tk \leq 1 
& 
\frac{\gamma-4-\tk(\gamma-2)}{2} < \tk^\prime < 1 - \frac{2(1+\tk)}{\gamma-2}
\\
{\rm or} & 
\\
\tk > 1
&
-\frac{\gamma-4+2\tk}{\gamma-2} < \tk^\prime < 1 - \frac{2(1+\tk)}{\gamma-2}
, 
\end{cases}
\label{eq: RPA boundaries}
\eea
and similarly for negative values of $\tk$. 
In the limit of $\gamma \to \infty$ this tends to 
Eq.~\eqref{eq: TG probability}, as one would expect. 
We tested our choice of probability regularisation in the 
TG limit by comparing its results with RPA and LL simulations for large values 
of $\gamma$ and we found good quantitative agreement (not shown). 

Using the new boundaries in 
Eq.~\eqref{eq: RPA boundaries}, 
the probability that a particle with wave vector 
$\tk$ scatters with the condensate 
in a time interval $\delta t$ 
(to any allowed $\tk^\prime$) remains finite for all allowed values of $\tk$. 
Namely, 
\bea
\int_{[\gamma-4-\tk(\gamma-2)]/2}^{1 - 2(1+\tk)/(\gamma-2)} \P_{k \to k^\prime} \: d\tk^\prime 
&=& 
\frac{\gamma^2}{\pi^3} \frac{\varepsilon_F}{\hbar} \delta t 
  \ln \left[ \frac{\gamma-2}{2} \right] 
\nonumber 
\eea
if $1-4/\gamma < \tk \leq 1$, and 
\bea
\int_{-(\gamma-4+2\tk)/(\gamma-2)}^{1 - 2(1+\tk)/(\gamma-2)} \P_{k \to k^\prime} \: d\tk^\prime 
&=&
\frac{\gamma^2}{\pi^3} \frac{\varepsilon_F}{\hbar} \delta t 
  \ln \left[ \frac{\tk+1-4/\gamma}{\tk-1+4/\gamma} \right] 
,
\nonumber
\eea
if $\tk > 1$. 
Notice that the maximum over $\tk>1$ of the 
logarithmic contribution in the second case is in fact the same as the 
first case: $\ln[(\gamma-2)/2]$. 
Our stochastic approach is therefore valid, provided that we choose 
\beq
\delta t \lesssim 
\frac{\pi^3}{\gamma^2} \frac{\hbar}{\varepsilon_F} 
  \left\{ \ln \left[ \frac{\gamma-2}{2} \right] \right\}^{-1}
. 
\eeq
For the typical system parametres considered in this work, the upper bound 
for $\delta t$ scales as $(\gamma^2 \ln \gamma)^{-1}$~milliseconds. 
This is satisfied for instance by choosing 
$\delta t \lesssim 0.01$~$\mu$s up to $\gamma = 100$. 

\begin{figure}[hb]
\includegraphics[width=0.95\columnwidth]
                {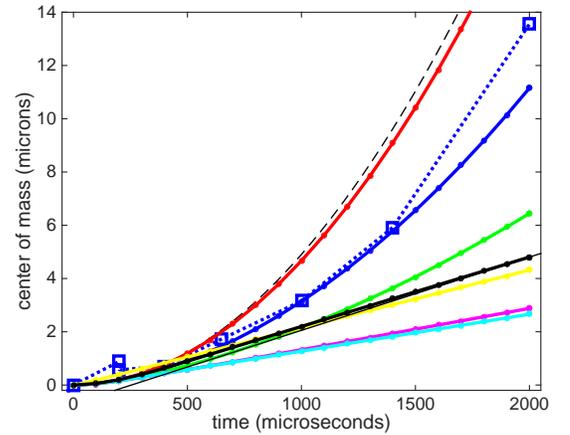}
\caption{
\label{fig: cm TG} 
(Colour online) --- 
Position of the center of mass as a function of time from our simulations 
in the TG limit, considering a uniform 1D gas with $n=1.2$~$\mu$m$^{-1}$ 
and $\gamma=4.1,\,5,\,6,\,8,\,10,\,25,\,100$ 
(red, blue, green, magenta, cyan, yellow, and black, respectively). 
The black dashed line represents the free-fall curve. 
The black solid line corresponds to the expected behaviour in the 
$\gamma \to \infty$ limit (i.e., terminal velocity $v_F$). 
Blue open squares (joined by a dotted line) represent the experimental data 
from Fig.~3 in Ref.~\onlinecite{Palzer2009}. 
}
\end{figure}

We can then implement our stochastic approach using the inverse 
transform sampling analytically in the TG limit. 
Fig.~\ref{fig: cm TG} shows the resulting behaviour of the centre of mass 
(CM) motion from TG 
simulations for a uniform 1D gas of density $n=1.2$~$\mu$m$^{-1}$ and 
$\gamma = 4.1,\,5,\,6,\,8,\,10,\,25,\,100$, to be contrasted with the results 
presented in the main text, Fig.~\ref{fig: cm vs expm}. 

We notice that the CM motion becomes asymptotically linear in time within the 
simulation time window for $\gamma \gtrsim 7$, 
suggesting that the falling atoms reach terminal velocity. 
The value of the terminal velocity is non-monotonic in $\gamma$: 
it initially decreases (in agreement with Ref.~\onlinecite{Rutherford2011}) 
with increasing $\gamma$, and later increases and tends asymptotically to 
$v_F$ in the $\gamma \to \infty$ limit, as expected. 

Reasonable agreement with the experimental results can only be obtained in 
the weak coupling limit ($\gamma \sim 5$), which is in contradiction with 
the TG limit (and even with the RPA approximation, which has a hard limit 
of applicability of $\gamma > 4$, and is known to begin to fit reasonably 
well the LL DSF only for $\gamma \gtrsim 10$~\cite{BrandCherny}). 
%
%

%
%


\begin{thebibliography}{99}

\bibitem{Palzer2009}
S.~Palzer, C.~Zipkes, C.~Sias, and M.~K\"{o}hl, 
\prl~\textbf{103}, 150601 (2009). 

\bibitem{Zipkes2010}
C. Zipkes, S. Palzer, C. Sias, and M. K\"{o}hl, 
Nature (London)~\textbf{464}, 388 (2010). 

\bibitem{Wicke2010}
P. Wicke, S. Whitlock, and N. J. van Druten, 
arXiv:1010.4545v1 (2010). 

\bibitem{Catani2012}
J. Catani, G. Lamporesi, D. Naik, M. Gring, M. Inguscio, F. Minardi, 
A. Kantian, and T. Giamarchi, 
\pra~\textbf{85}, 023623 (2012). 

\bibitem{Fukuhara2013}
T. Fukuhara, A. Kantian, M. Endres, M. Cheneau, P. Schauss, S. Hild, 
D. Bellem, U. Schollw\:{o}ck, T. Giamarchi, C. Gross, I. Bloch, and S. Kuhr, 
Nat. Phys.~\textbf{9}, 235 (2013). 

\bibitem{Hakim1997}
V. Hakim, 
\pre~\textbf{55}, 2835 (1997). 

\bibitem{Zvonarev2007}
M. B. Zvonarev, V. V. Cheianov, and T. Giamarchi, 
\prl~\textbf{99}, 240404 (2007). 

\bibitem{Astrakharchik2004}
G. E. Astrakharchik and L. P. Pitaevskii, 
\pra~\textbf{70}, 013608 (2004). 

\bibitem{Ponomarev2006}
A. V. Ponomarev, J. Madro\~{n}ero, A. R. Kolovsky, and A. Buchleitner, 
\prl~\textbf{96}, 050404 (2006). 

\bibitem{Girardeau2009}
M. D. Girardeau and A. Minguzzi, 
\pra~\textbf{79}, 033610 (2009). 

\bibitem{Sykes2009}
A. G. Sykes, M. J. Davis, and D. C. Roberts, 
\prl~\textbf{103}, 085302 (2009). 

\bibitem{Cherny2009}
A. Yu. Cherny, J.-S. Caux, and J. Brand, 
\pra~\textbf{80}, 043604 (2009). 

\bibitem{Gangardt2009}
D. M. Gangardt and A. Kamenev, 
\prl~\textbf{102}, 070402 (2009). 

\bibitem{Johnson2011}
T. H. Johnson, S. R. Clark, M. Bruderer, and D. Jaksch, 
\pra~\textbf{84}, 023617 (2011). 

\bibitem{Rutherford2011}
L.~Rutherford, J.~Goold, Th.~Busch, and J.~F.~McCann, 
\pra~\textbf{83}, 055601 (2011). 

\bibitem{Goold2011}
J. Goold, T. Fogarty, N. Lo Gullo, M. Paternostro, and Th. Busch, 
\pra~\textbf{84}, 063632 (2011). 

\bibitem{Schecter2012}
M. Schecter, A. Kamenev, D. M. Gangardt, and A. Lamacraft, 
\prl~\textbf{108}, 207001 (2012). 

\bibitem{Bonart2012}
J.~Bonart and L.~F.~Cugliandolo, 
Europhys.~Lett.~\textbf{101}, 16003 (2013); 
\pra~\textbf{86}, 023636 (2012). 

\bibitem{Mathy2012}
C. J. M. Mathy, M. B. Zvonarev, and E. Demler, 
Nat.~Phys.~\textbf{8}, 881 (2012). 

\bibitem{Knap2013}
M. Knap, C. J. M. Mathy, M. Ganahl, M. B. Zvonarev, and E. Demler, 
Phys.~Rev.~Lett.~\textbf{112}, 015302 (2012). 

\bibitem{Girardeau1960}
M.~D.~Girardeau, 
J. Math. Phys.~\textbf{1}, 516 (1960). 

\bibitem{Cherny2012}
A. Yu. Cherny, J.-S. Caux, and J. Brand, 
Front.~Phys.~\textbf{7}, 54 (2012). 

\bibitem{footnote: approach}
In support of our choice of approach, we point out that 
Ref.~\onlinecite{Johnson2011} shows how a perturbative approach to study 
these systems appears to be valid beyond the perturbative regime. 
Moreover, both Ref.~\onlinecite{Ponomarev2006} and 
Ref.~\onlinecite{Johnson2011} demonstrate the appropriateness and 
effectiveness of a (semi-classical) stochastic description. 

\bibitem{Bogoliubov1947}
N.~N.~Bogoliubov, 
J.~Phys.~(USSR)~\textbf{11}, 23 (1947), 
reprinted in The Many-body Problem, edited by D.~Pines 
(Benjamin, New York, 1961).

\bibitem{Caux} 
J.-S. Caux and P. Calabrese, Phys. Rev. A 74, 031605 (2006).

\bibitem{SiasKoehl}
C.~Sias and M.~K\"{o}hl, 
(private communication). 

\bibitem{footnote: 1D gas profile}
For the relatively small values of $\gamma$ that produce the best agreement 
with the experiments according to our simulations, we find that the exact 
DSF from LL calculations yields results similar to the DSF from the BG 
approximation. 
Within the BG approximation, we were then able to test the effects of a 
\textit{static} non-uniform 1D gas density profile on our results and 
demonstrate that they are negligible 
(see Appendix~\ref{app: 1D gas profile dep}). 

\bibitem{footnote: smoothing}
Smoothing of the Wigner function is required (over appropriately 
wide regions) to ensure non-negativity, as we want to interpret it as 
a probability distribution function (see Ref.~\onlinecite{Cartwright1976}). 
The typical approaches are a sliding average over a constant interval, 
or a Gaussian convolution. We chose the latter for numerical 
tractability, as the result can be conveniently written in terms of 
error functions, Eq.~\eqref{eq:Gfinal}. 

\bibitem{Cartwright1976}
N. D. Cartwright, 
Physica~\textbf{83A}, 210 (1976). 

\bibitem{footnote: expm corrections background}
In the experimental data for $\gamma = 1.1$ and $\gamma = 1.9$, 
we subtracted a background ($0.0046$ and $0.023$, respectively, in 
optical intensity units) so that the right tail of the position distribution 
drops down to approximately zero. This is justified by the fact that such 
a background ought to be spurious, as 
it could only arise from atoms accelerating faster than $g = 9.813$~m/s$^2$ 
(unphysical). 

\bibitem{footnote: expm corrections long time}
We note that the time reported in Ref.~\onlinecite{Palzer2009}, $18.7$~ms, 
appears to be inconsistent with the position of the main peak. 
Assuming $g=9.813$~m/s$^2$, the main peak of free falling atoms starting at 
$t=0$ in a gaussian packet centerd at $x=0$ should appear at 
$x=1,716$~$\mu$m after $18.7$~ms. We find instead that a time of 
$18.6$~ms is in better agreement with the data and we used it in our 
simulations. 
This is consistent with the fact that the falling packet is created with a 
laser pulse of finite duration. Indeed, comparing the non-interacting data 
from Ref.~\onlinecite{Rutherford2011} with the analytical free-fall 
behaviour $g (t - t_0)^2 / 2$, we find that a shift $t_0 \simeq 90$~$\mu$s 
is needed to achieve good agreement, which is consistent with a difference 
between $18.6$~ms and $18.7$~ms. 

\bibitem{BrandCherny}
J.~Brand and A.~Y.~Cherny, 
\pra~\textbf{72}, 033619 (2005); 
A.~Y.~Cherny and J.~Brand, 
\pra~\textbf{73}, 023612 (2006). 

\end{thebibliography}
\end{document}